# Potential customer mining application of smart home products based on  LightGBM+PU learning and Spark ML algorithm practice

Duan Zhihua     Wang JiaLin

China Telecom Shanghai Branch     Top AI lab in Silicon Valley

**Abstract**: This paper studies the case of big data-based intelligent product potential customer mining internal competition in China Telecom Shanghai Company. Huge amounts of data based on big data table, the use of machine Learning and data analysis technology, using the algorithm of LightGBM, PySpark machine Learning algorithms, Positive Unlabeled Learning algorithm, and predict whether customers buy whole house wifi product and BoBo BaoHe、TianYi ShenXue products, precision marketing into artificial intelligence for the customer, large data capacity, promote the development of intelligent products of the company.

## 1 Background

This paper is a case of the internal competition of China Telecom Shanghai to explore potential customers of smart products based on big data. The internal competition includes preliminary contest and final contest. The competition data is provided by the Big data table of Shanghai Company. Preliminary contest based on the customer identification and customer call data, predict whether customers buy whole house wifi products,1 means to buy, 0 means not to buy.Final contest Based on a large amount of unmarked customer identification and customer call data, it is estimated that 10,000 users are likely to buy BoBo BaoHe and Tianyi Shenxue products.

Data characteristics of preliminary contest data set:1. Large amount of training data: customer identification data (about 2.5 million records, 129 fields), customer call data (about 1.3 billion records, 86 fields), training set (80,000 records), test set (50,000 records). 2. Data sparsity: some features do not have values, and the number of some associated records is small. 3. Some features are strongly correlated with the training set, resulting in overfitting.

The data characteristics of the final contest data set: 1. There are only a small number of positive sample data. The competition official only provided 1000 users' data that have purchased BoBo BaoHe and 500 users' data that have purchased Tianyi Shenxue. 2. Provided a huge amount of unmarked training data, including 180 million customer identification records and 24.3 billion customer call records. 3. Based on a small number of positive samples and a large number of unmarked samples, each predicted 10,000 potential customers list.

In this paper, LightGBM algorithm was applied in the preliminary competition. Meanwhile, PySpark machine learning algorithm was used to predict whether customers would buy whole-house wifi products and submit the optimal prediction results. The main contents of the preliminary contest are as follows: machine learning data preprocessing, data exploration, correlation analysis, LightGBM model analysis, feature derivation and model optimization; PySpark is used to study algorithm models based on distributed cluster, such as Naive Bayes, decision tree, random forest and logistic regression.

 In this paper, PU Learning machine Learning algorithm was adopted in the final competition.

Based on a very small number of Positive samples and a large number of Unlabeled samples of large scale, LightGBM algorithm was used for model training, and 10,000 users of BoBo BaoHe and Tianyi Shenxue products were predicted.

## 2 LightGBM algorithm application in preliminary contest

LightGBM is a gradient enhancement framework based on tree learning algorithm provided by Microsoft. LightGBM has the following advantages: fast training speed, high efficiency, low memory utilization, accurate prediction, support for parallel and GPU learning, and ability to process large-scale data. LightGBM is widely used in Various Kaggle machine learning competitions.

In the preliminary contest, feature engineering technologies such as data preprocessing, data exploration, correlation analysis, model analysis, feature derivation and model optimization commonly used in the competition were adopted. LightGBM algorithm was used as the baseline model code to realize the prediction of whole-house wifi products, and the classical dichotomy in machine learning was explored in practice.

### 2.1 Data preprocessing

Customer identification data, customer call data, training set and test set are placed in the internal big data workbench. Data is stored based on Hadoop big Data platform and statistical analysis is carried out with Hive tool. For example, the three tables of customer identification data, customer call data and training set are directly joined, because of the large amount of data and MapReduce runtime may get stuck, therefore, the customer identification table according to the device number after polymerization import new customer identification table (about 130000 records), will be customer data according to the device number polymerization import new customer data table (about 23 million records), and then respectively left joined with training set and test set, export data is saved as a new training set (80000 records), the test set (50000 records).

### 2.2 Data exploration

Associated after new training set and test set including asset state name, package name, the name of the sales strategy, access to the Internet (interest), and preferences terminal brand, monthly ARPU, vice card number, the terminal type, customer address customer monthly ARPU, access time (province), the application name, visits, visit the URL information fields, such as through visualization, computational statistics to explore the training set and testing set the overall distribution of information, to find possible outlier data record, further familiar with the business data.

In the case of the preliminary competition, the distribution of the values of each field is counted. Convert business time, extract year, month, day, week, quarter and other information from network access time, customer's birthday, expiration date of agreement constraint, and opening date time, and sum up, calculate average, and deal with outliers for application access times of customers; Analyze the type of each field in the data set (string type, numeric type); Specify category characteristics, and visualize the distribution difference of different categories for whether to buy wifi products; The mean value, extreme value and standard deviation were used to analyze the descriptive statistics. In EDA visualization data exploration, for the feature of monthly ARPU of address customers, based on the sample size of the data set, there is little difference in the distribution of training set and test set.

**2.3 Correlation analysis**

The number of features in the case data set of the preliminary competition is large, and the correlation analysis is based on the full amount of data to study the correlation between two or more features. If there are many missing values in the two features, the calculated correlation deviation is large. Therefore, the advanced null value search is used to count the field groups with the same null value row record number, and the missing value features are put into a group to observe the data distribution. Then, use the correlation heat map visualization, agreement date field to extract the "year, month, day, week, one day a year, some weeks a year, quarterly" field information with the same lack of record, the characteristic of high correlation ($r > 0.97$) was divided into a group, according to 0.97 threshold time characteristics, and then select the different data values in each group with more columns to take the place of this group, eliminate redundant fields, this set of selected features for "year, day, week, year one day, a year a week". And so on, the correlation of each group of features is calculated to obtain the required feature set.

**2.4 Build LightGBM baseline model and model analysis**

According to correlation calculation, select required feature columns from training set and test set, set LightGBM super parameters, construct LightGBM baseline model for training and prediction, and the prediction results show that the baseline model appears overfitting phenomenon. A fitting because of the differences between training set and testing set data distribution, This case used against method validation data set offset, the distribution of the training set of tags is set to 1, the test set of tags is set to 0, training a LightGBM model, through the AUC judge training set and testing set data distribution difference degree, and through the calculation of characteristic importance, determine which features brought a larger deviation. It was found that the characteristics of DEV_BUREAU_NAME had a strong correlation with the training set, so the characteristics of the bureau name were deleted, and the LightGBM baseline model was used for cross-validation, and the predicted score was significantly improved.

Using antagonistic validation method can find the fitting part features, but the importance of features of the ranking of the top off one by one do, discover prediction assessment score is greatly reduced, so you can't completely dependent on the antagonistic verification method, case and Kris verification method for each feature independently cross validation, if the characteristics of the training set AUC contribution is small, the validation set AUC contribution is less than 0.5, just get rid of this characteristic, on the basis of the continued to do importance characteristic analysis, fine-tuning, choice of feature set prediction score and improved.

**2.5 Feature derivation and model optimization**

Feature derivation is based on the combination of existing features to generate new features. This case choose important features, such as average monthly ARPU, address customer assets structure, customer monthly ARPU characteristics, such as the important category of value distribution of encoding, as well as the important category of two merging, a new feature of two categories, or three three merger, the three category features of a new, training LightGBM model to forecast, score rise around 3.

On this basis, model optimization was carried out, feature crossing was carried out according to the importance of features, GroupBy aggregation was conducted between categories and continuous values, and 180 new features were generated. Again, the required features were selected according to the previous feature selection steps, and then LightGBM model and prediction were trained to increase the prediction score by about 5.

**2.6. Machine learning algorithms based on distributed PySpark**

ML and MLlib machine learning libraries are provided in Spark to make distributed machine learning extensible and easy. At a higher level, Spark provides the following tools, ML algorithms: common learning algorithms such as classification, regression, clustering, and collaborative filtering; Characterization: feature extraction, transformation, reduction and selection; Pipes: Tools for building, evaluating, and adjusting ML pipes; Persistence: Saves and loads algorithms, models, and pipes; Utilities: Linear algebra, statistics, data processing, etc.

Preliminary case Spark big data clustering based on distributed machine learning field, from PySpark machine learning characteristics of the engineering library（pyspark.ml.feature）import HashingTF、IDF, TF represents the number of occurrences of words, IDF represents the inverse word frequency of the word, using TF/IDF tools for customers to access application name, customer access to the host URL constructing text characteristic, The VectorAssembler tool of Pyspark was used to integrate text features, numerical features, category features and time features into feature vectors, and Pyspark machine learning models (naive Bayes, decision tree, random forest, logistic regression) were used for training and prediction to predict whether customers would purchase whole-house wifi products.

**2.7 Preliminary contest model prediction results**

LightGBM、Naive Bayes、decision tree、random forest and Logistic regression models were built to generate the prediction results of the test set in the specified equipment order and submit them to the official website platform for scoring. F1 was adopted as the scoring standard in this competition, and the scores of each model were as follows:

| Algorithm model | score |
| --- | --- |
| LightGBM | 350559.4 |
| Naive Bayes | 329393.22 |
| decision tree | 318355.36 |
| random forest | 343471.21 |
| Logistic regression | 339417.16 |
| OneVsRest | 339417.16 |

Table 1 Scores of each model

**3 final contest model prediction**

Based on massive customer identification and customer call data, the final contest is expected to buy BoBo BaoHe and Tianyi Shenxue products.

**3. 1 Pretreatment of massive unlabeled data**

The training data of the finals contest provides a huge amount of unmarked training data, including 180 million customer identification records and 24.3 customer call records. The training data is stored on the Hadoop big Data platform. Article 24.3 billion the record has been partitioned according to day, hour, to extract the text features of each field, such as access to the URL text length, number of visits, the device number, etc., according to the number in the Hive SQL polymerization export data every day, and then inserted into the new customer identification in the table, generate new table record number approximately 10 million records, and then with the customer data, BoBo billed BaoHe and Tianyi Shenxue data association, respectively to generate BoBo BaoHe, Tianyi Shenxue training data, BoBo BaoHe includes 1000 positive samples, Tianyi Shenxue includes 500 positive samples, and the others are massive unmarked records.

### 3. 2 PU algorithm Implementation

In the practical business scenarios of BoBo BaoHe and Tianyi Shenxue, only some positive samples bought by BoBo BaoHe and Tianyi Shenxue are provided, while the others are massive unmarked samples. These unmarked samples may be positive samples or negative samples. How do you label an unlabeled sample? This is the content of this final contest research PU algorithm, based on a small number of positive samples and a large number of unmarked samples for classification prediction.

The PU algorithm implemented in the final contest is as follows:

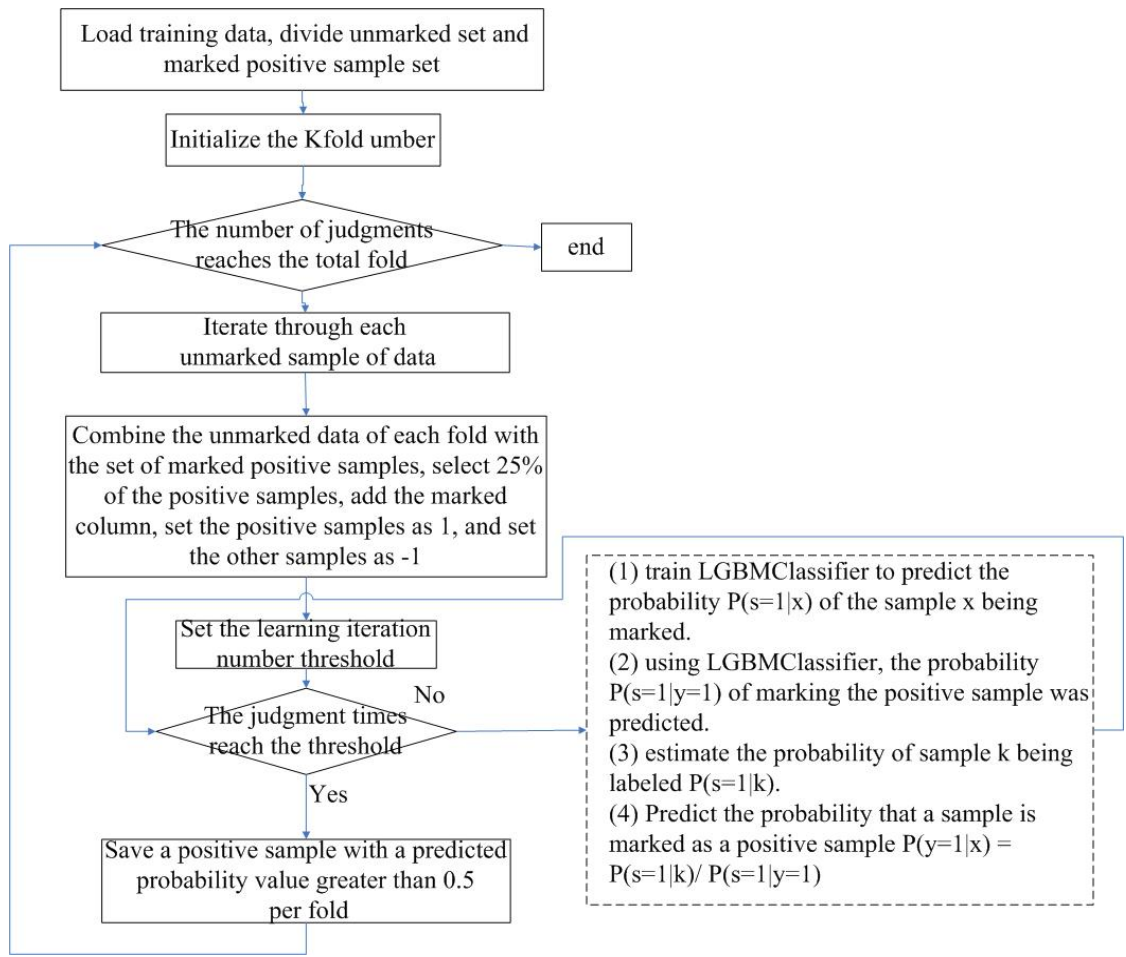

Figure 1 PU algorithm

### 3.3 The forecast result of the final contest

From 180 million customer identification records and 24.3 billion customer call records, based on 1000 ordered BoBo BaoHe and 500 ordered Tianyi Shenxue, the modeling design was carried out using LGBMClassifier to achieve the successful prediction of 10,000 potential customer lists in the case of only a small number of positive samples and a large number of unmarked samples.

**4 Conclusion and Prospect**

In this paper, internal competition for potential customers of smart products based on big data was studied. Further studies were performed on LightGBM algorithm, PySpark machine Learning algorithm and Positive Unlabeled Learning algorithm.

Looking ahead, Spark + AI 2020 calls for further exploration of how Spark and AI together shape the world of cognitive computing and how ARTIFICIAL intelligence can create new opportunities in business through innovative use cases. The Spark + AI 2020 Summit will focus on all aspects of AI, from autonomous driving to voice and image recognition, from intelligent chatbots and new deep learning frameworks to efficient machine learning algorithms, models and methods, including vision, voice, deep learning and large-scale distributed learning. Among them, NLP will become an important development field of artificial intelligence, and the research on NLP will be full of opportunities and challenges.

**5 Thank**

Thanks to China Telecom Shanghai For organizing the "Wing code first" software talent competition, which provides massive data based on the internal big data workbench and uses the tools and capabilities of the big data workbench for data preprocessing and modeling development. Each participant successfully completed the competition based on the online programming environment.